\documentclass[preprint,12pt,sort&compress,nopreprintline]{elsarticle}

\makeatletter
  \long\def\pprintMaketitle{\clearpage
  \iflongmktitle\if@twocolumn\let\columnwidth=\textwidth\fi\fi
  \resetTitleCounters
  \def\baselinestretch{1}%
  \printFirstPageNotes
  \begin{center}%
 \thispagestyle{pprintTitle}%
   \def\baselinestretch{1}%
    \Large\@title\par\vskip18pt
    \normalsize\elsauthors\par\vskip10pt
    \footnotesize\itshape\elsaddress\par\vskip10pt
    
    \hrule\vskip12pt
    \ifvoid\absbox\else\unvbox\absbox\par\vskip10pt\fi
    \ifvoid\keybox\else\unvbox\keybox\par\vskip10pt\fi
    \hrule\vskip12pt
  \end{center}%
  \gdef\thefootnote{\arabic{footnote}}%
}
\makeatother

\usepackage{siunitx}
\usepackage{amsmath}
\usepackage{graphicx}
\usepackage{float}
\usepackage{subfigure}
\usepackage{epstopdf}
\usepackage{color}
\usepackage{multirow}
\usepackage{setspace}
\usepackage{overpic}
\usepackage{amssymb}
\usepackage{mathrsfs}
\usepackage[colorlinks,urlcolor=blue, citecolor=blue,linkcolor=blue] {hyperref}
\usepackage{lineno}
\usepackage{bm}
\usepackage{rotating}
\usepackage{array}
\usepackage{booktabs}
\usepackage{besphysics}
\usepackage[utf8]{inputenc}
\usepackage{enumitem}
\newcommand{\BESIIIorcid}[1]{\href{https://orcid.org/#1}{\hspace*{0.1em}\raisebox{-0.45ex}{\includegraphics[width=1em]{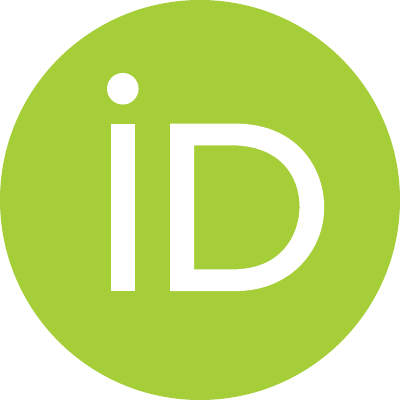}}}}

\let\oldequation\equation
\let\oldendequation\endequation
\renewenvironment{equation}
 {\linenomathNonumbers\oldequation}
 {\oldendequation\endlinenomath}
 
\begin{document}
\journal{Science Bulletin}

\begin{frontmatter}

\title{\boldmath First determination of the full spin density matrix in $\psi(3686)\to\Lambda\bar{\Lambda}$}

\author{BESIII Collaboration\footnote{Corresponding author (email: besiii-publications@ihep.ac.cn)}\footnote{Authors are listed at the end of this paper.}}

\begin{abstract}
Using $(448.1\pm2.9)\times10^{6}$ $\psi(3686)$ events collected with the BESIII detector, 
we present the first complete determination of the spin density matrix in $\psi(3686)\to\Lambda\bar{\Lambda}$. This provides a unified framework to study both the production dynamics and the quantum correlations of the $\Lambda\bar{\Lambda}$ system. From a five-dimensional joint angular analysis, we determine $\alpha_{\psi}=0.82\pm0.02_{\mathrm{stat.}}\pm0.01_{\mathrm{syst.}}$ and $\Delta\Phi=(20.6\pm3.6_{\mathrm{stat.}}\pm0.6_{\mathrm{syst.}})^{\circ}$.
The non-zero phase implies a transverse polarization of $\Lambda$, with a maximum value of $P_{y} = 0.074 \pm 0.013_{\mathrm{stat.}} \pm 0.003_{\mathrm{syst.}}$ and a significance of 6.3$\sigma$. 
The same framework also gives access to the spin correlations of the baryon pair and enables tests of quantum nonclassicality.
The average of the Bell nonlocality measure is determined to be $\bar{m}_{12} = 1.266 \pm 0.078_{\mathrm{stat.}} \pm 0.049_{\mathrm{syst.}}$, yielding a violation of the  Clauser-Horne-Shimony-Holt inequality with a significance of 3.7$\sigma$.
These results establish $\psi(3686)\to\Lambda\bar{\Lambda}$ as a unique laboratory for unified studies of hadron production dynamics and quantum entanglement in a system of massive baryons.
\end{abstract}


\end{frontmatter}


 
\section{Introduction}\label{sec:intro}

Spin polarization in high-energy collisions offers critical insights into Quantum Chromodynamics (QCD)~\cite{Barone:2001sp}. Hyperons, spin-$1/2$ baryons containing strange quarks, are especially powerful probes due to their self-analyzing weak decays, which enable direct access to their spin orientation. Polarization of hyperons has been observed in hadronic collisions since the 1970s~\cite{Bunce:1976yb,Abramowicz:1975fx,Heller:1978ty,Lomanno:1979udv}. 
Such observations triggered extensive studies of polarization phenomena in QCD. More recently, hyperon polarization has been measured in heavy-ion and proton-proton collisions~\cite{ALICE:2019onw,STAR:2017ckg,ATLAS:2014ona}. However, exclusive electron-positron ($e^+e^-$) collisions provide a much cleaner environment to access fundamental features of baryon production. Such a clean experimental environment at BESIII also enables precision studies of heavy hadron spectroscopy and sensitive searches for rare charmonium decays~\cite{LHCb:2025mnp,BESIII:2024axd,BESIII:2025tmo}. In exclusive $e^+e^-$ collisions, hyperon-antihyperon pair production is governed by electromagnetic form factors (EMFFs), typically the electric ($G_\mathrm{E}$) and magnetic ($G_{\mathrm{M}}$) form factors, which encode the internal structure of baryons~\cite{Dubnickova:1992ii,Cabibbo:1961sz,Brodsky:2003gs}. The form factors govern the polar angular distribution, $1 + \alpha \cos^2\theta_{\Lambda}$, and induce transverse polarization via their relative phase $\Delta\Phi$, making $\alpha$ and $\Delta\Phi$ key observables of the production dynamics.

Although the EMFF framework is well-established, several key issues remain unresolved.  Earlier studies of $e^+e^- \to \Lambda\bar{\Lambda}$ were carried out by the BaBar experiment using initial-state radiation, where the $\Lambda$ transverse polarization was found to be consistent with zero within relatively large statistical uncertainties~\cite{BaBar:2007fsu}. To date, the energy dependence of $\Lambda$ polarization is still poorly understood, with high precision achieved only at the $J/\psi$ resonance~\cite{BESIII:2018cnd}. A further puzzle is the contrasting behaviors of the $\Lambda$ and $\Sigma^0$ hyperons. The angular distribution parameter $\alpha_{\psi}$ is positive for $\Lambda\bar{\Lambda}$~\cite{BES:2005qcn} but negative for $\Sigma^0 \bar{\Sigma}^0$~\cite{BESIII:2024nif} in the $J/\psi$ decays, despite their identical valence quark content. Proposed explanations include SU(3)-flavor symmetry breaking and diquark configurations~\cite{Alekseev:2018qjg,BaldiniFerroli:2019abd,Ferroli:2020mra,Jaffe:2003sg}. The CLEO-c Experiment has presented evidence for the effects of diquark correlations~\cite{Dobbs:2017hyd}, suggesting that the charmonium states provide complementary information on the production dynamics and internal structures of hyperons. The complexity deepens at the $\psi(3686)$, where both the hyperon polarization and $\alpha_{\psi}$ for $\Sigma^0\bar{\Sigma}^0$ flip sign relative to $J/\psi$~\cite{BESIII:2024nif}. A similar polarization behavior has also been observed for $\Sigma^+$ hyperons~\cite{BESIII:2025jxt}.  Whether $\Lambda$ shows a similar behavior at the $\psi(3686)$ remains unknown, since no $\Lambda$ polarization measurement in $\psi(3686) \to \Lambda\bar{\Lambda}$ has been reported. Addressing this gap is essential for a unified understanding of hyperon production across the charmonium family~\cite{Bigi:2017eni,BESIII:2020nme}.

These open questions call for a more complete description of hyperon pair production. Such a description is naturally provided by the spin density matrix, which unifies production dynamics and hyperon–antihyperon spin correlations within a single framework. Importantly, it also provides direct access to quantum entanglement between the two hyperons. While quantum entanglement has been extensively tested in photons, atoms, and solid-state systems~\cite{Aspect:1982fx, Hensen:2015ccp}, experimental studies involving strongly interacting massive particles remain scarce. Hyperon–antihyperon pairs produced in $e^+e^-$ annihilation form a naturally entangled system, where parity-violating weak decays enable event-by-event reconstruction of the spin density matrix, allowing quantitative studies of entanglement and tests of Bell-type inequalities~\cite{Tornqvist:1980af}. The utility of such systems has recently been demonstrated using entangled $\Xi^0\bar{\Xi}^0$ pairs, which enabled a precise study of the weak radiative decay $\Xi^0\to\Lambda\gamma$ and its decay
asymmetry~\cite{BESIII:2024lio}.

In this work, we present the first complete determination of the spin density matrix in $\psi(3686)\rightarrow\Lambda\bar{\Lambda}$ using a five-dimensional joint angular analysis, based on the same data sample as in the previous analysis~\cite{BESIII:2017kqw}. Unlike the previous analyses based on partially integrated angular distributions, this approach gives direct access to the full spin structure of the $\Lambda\bar{\Lambda}$ system. At vector charmonium resonances, baryon--antibaryon production is predominantly governed by strong interactions via three-gluon annihilation, with the corresponding production amplitudes encoded in the psionic form factors $G_\mathrm{E}^{\psi}$ and $G_{\mathrm{M}}^{\psi}$~\cite{Faldt:2017kgy, Faldt:2013gka, Faldt:2016qee}. The extracted parameters $\alpha_{\psi}$ and $\Delta\Phi$ directly determine the magnitude ratio and relative phase of these form factors. Building on this unified framework, the spin density matrix enables a quantitative study of quantum entanglement in the $\Lambda\bar{\Lambda}$ system by testing the Clauser-Horne-Shimony-Holt (CHSH) inequality and quantifying entanglement via Wootters' concurrence~\cite{Wu:2024asu}. In this way, $\psi(3686)\to\Lambda\bar{\Lambda}$ provides a unified experimental framework for studying hadron production dynamics and quantum nonclassicality in a system of massive baryons.

\section{Methods}\label{sec:methods}

\begin{figure}[hbtp]
    \centering
    \includegraphics[width=0.65\linewidth,trim=15 25 20 10,angle=0]{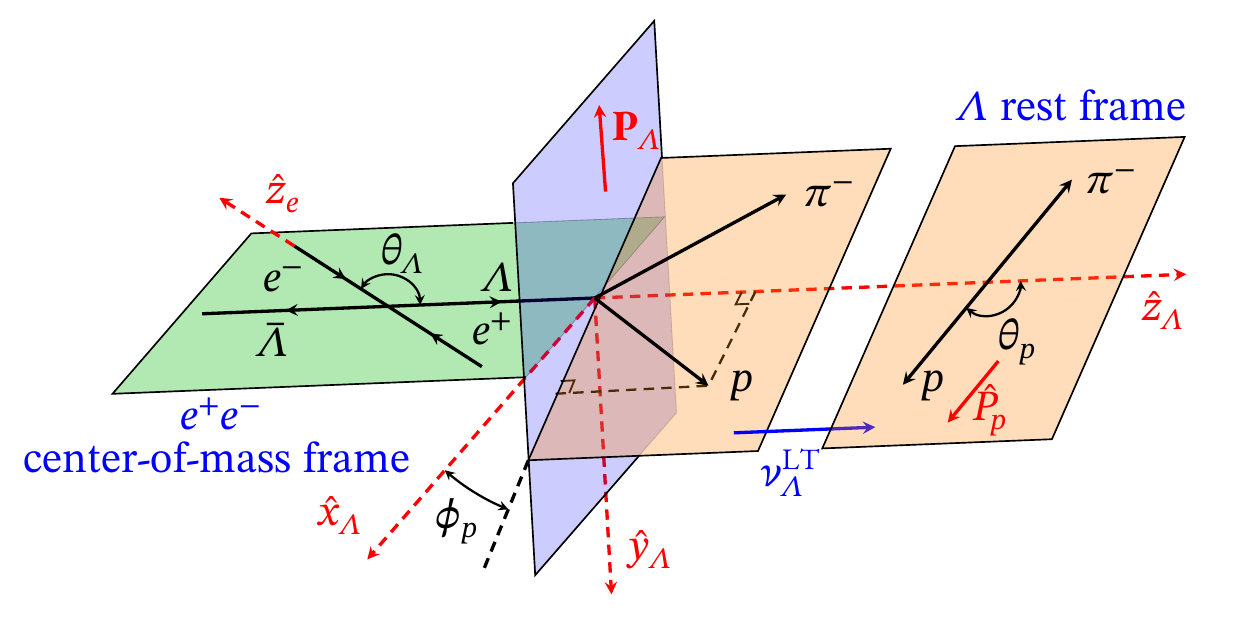}
\caption{Coordinate system for the reaction $e^+e^-\to\psi(3686)\to\Lambda\bar{\Lambda}\to p \pi^-\bar{p}\pi^+$. The $\hat{z}_{e}$ axis is defined along the direction of the $e^+$ beam. The $\hat{z}_{\Lambda/\bar{\Lambda}}$ axis is defined along the $\Lambda/\bar{\Lambda}$ momentum direction. The axis $\hat{y}_{\Lambda/\bar{\Lambda}}=\hat{z}_{e}\times\hat{z}_{\Lambda/\bar{\Lambda}}$ is normal to the scattering plane and $\hat{x}_{\Lambda/\bar{\Lambda}}=\hat{y}_{\Lambda/\bar{\Lambda}}\times\hat{z}_{\Lambda/\bar{\Lambda}}$. The vector $\mathbf{P}_{\Lambda}$ is the hyperon polarization vector and $\mathbf{\hat{p}}_{p}$ is the unit vector along the momentum of the proton in the $\Lambda$ rest frame. The angle $\theta_{\Lambda}$ is the angle between the  $\Lambda$ momentum and the $e^+$ beam direction in the $e^+e^-$ center-of-mass frame; $\theta_{p}$ and $\phi_{p}$ are the proton helicity angles from the $\Lambda\to p\pi^-$ decay in the $\Lambda$ rest frame, $\theta_p$ is the polar angle of the proton momentum with respect to $\hat z_\Lambda$, and $\phi_p$ is the corresponding azimuthal angle around $\hat z_\Lambda$, measured from $\hat x_\Lambda$ toward $\hat y_\Lambda$. The $\nu^{\mathrm{LT}}_{\Lambda/\bar{\Lambda}}$ is the Lorentz transformation with the velocity of $\nu_{\Lambda/\bar{\Lambda}}$.}
\label{fig1}
\end{figure}

To realize this unified framework experimentally, we perform a full angular analysis of the production and decay chain $e^{+}e^{-}\to \psi(3686) \to \Lambda( p\pi^{-}) \bar{\Lambda}(\bar{p}\pi^{+})$.
The differential cross-section is described with five observables $\boldsymbol{\xi}= (\theta_{\Lambda}, \theta_{p}, \phi_{p}, \theta_{\bar{p}}, \phi_{\bar{p}})$. The coordinate system and these angles are defined following the formalism of Ref.~\cite{Perotti:2018wxm} and illustrated in Fig.~\ref{fig1}. This full angular distribution is determined by a set of four parameters, $\boldsymbol{\Omega}=(\alpha_{\psi},\Delta\Phi,\alpha_{\Lambda},\alpha_{\bar{\Lambda}})$, which are independent of the $\Lambda$ scattering angle. The distribution can be written in a modular form as:
\vspace{-\parskip}
\begin{equation}
\mathcal{W}(\boldsymbol{\xi};\boldsymbol{\Omega}) = \sum_{\mu,\nu=0}^{3} \rho_{\mu\nu} a_{\mu 0}^{\Lambda} a_{\nu 0}^{\bar{\Lambda}}.
\label{eq:anglW}
\end{equation}
Here, $\rho_{\mu\nu}(\theta_{\Lambda};\alpha_{\psi},\Delta\Phi)$ is a $4\times4$ spin density matrix that characterizes the spin configuration of the entangled hyperon–antihyperon pair. The matrices $a_{\mu0}^{\Lambda}(1,\alpha_\Lambda\sin\theta_p\cos\phi_p,\alpha_\Lambda\sin\theta_p\sin\phi_p,\alpha_\Lambda\cos\theta_p )$ and $a_{\nu0}^{\bar{\Lambda}}
(1,\alpha_{\bar{\Lambda}}\sin\theta_{\bar{p}}\cos\phi_{\bar{p}},\allowbreak \alpha_{\bar{\Lambda}}\sin\theta_{\bar{p}}\sin\phi_{\bar{p}},\alpha_{\bar{\Lambda}}\cos\theta_{\bar{p}})$ describe the spin transfer in the parity-violating weak decays of $\Lambda$ and $\bar{\Lambda}$, respectively, relating the hyperon spin states to the proton and antiproton angular distributions. The production parameters $\alpha_{\psi}$ and $\Delta\Phi$, encoded in $\rho_{\mu\nu}$, govern the transverse polarization $P_{y}$ of the produced $\Lambda$ as well as the spin correlations $C_{ij}$. The matrix elements are related to $P_{y}(\theta_\Lambda)$ and $C_{ij}(\theta_\Lambda)$ in the following way:
\begin{equation}
\rho_{\mu\nu} = (1 + \alpha_{\psi} \cos^2\theta_\Lambda)
\begin{pmatrix}
1 & 0 & P_y & 0 \\
0 & C_{xx} & 0 & C_{xz} \\
-P_y & 0 & C_{yy} & 0 \\
0 & -C_{xz} & 0 & C_{zz}
\end{pmatrix}.
\label{eq:Cmatrix}
\end{equation}
For a given set of global parameters $\alpha_{\psi}$ and $\Delta\Phi$, the magnitude of $P_y$ and  the correlations $C_{ij}$ vary with $\theta_\Lambda$.  The nonzero components are given by
\begin{equation}
P_{y}(\theta_\Lambda) = \frac{\sqrt{1-\alpha_{\psi}^2}\sin\theta_{\Lambda} \cos\theta_{\Lambda}}{1+\alpha_{\psi} \cos^2\theta_{\Lambda}}\sin\Delta\Phi,
\label{eq:py}
\end{equation}
\begin{equation}
\begin{gathered}
C_{xx}(\theta_\Lambda) = \frac{\sin^{2}\theta_\Lambda} {1+\alpha_{\psi}\cos^{2}\theta_\Lambda}, \quad C_{yy}(\theta_\Lambda) = \frac{-\alpha_{\psi}\sin^{2}\theta_\Lambda}
{1+\alpha_{\psi}\cos^{2}\theta_\Lambda}, \quad C_{zz}(\theta_\Lambda) = \frac{\alpha_{\psi}+\cos^{2}\theta_\Lambda}{1+\alpha_{\psi}\cos^{2}\theta_\Lambda}, \\
C_{xz}(\theta_\Lambda)=C_{zx}(\theta_\Lambda) = \frac{\sqrt{1-\alpha_{\psi}^{2}}\,
\cos\Delta\Phi\, \sin\theta_\Lambda\cos\theta_\Lambda}
{1+\alpha_{\psi}\cos^{2}\theta_\Lambda}.
\end{gathered}
\label{eq:cij}
\end{equation}
This structure allows us to first extract the parameters $\alpha_{\psi}$ and $\Delta\Phi$ from the full data sample, and subsequently determine the specific values of the spin correlation elements $C_{xx} , ..., C_{zz}$ by performing independent fits in separate $\cos\theta_{\Lambda}$ bins. In this way, the same spin density matrix formalism provides the common basis for the measurements of production observables, spin correlations, and the subsequent tests of Bell nonlocality and entanglement.

\section{Data analysis}
The full process $e^+e^-\to\psi(3686)\to\Lambda\bar{\Lambda}\to p \pi^-\bar{p}\pi^+$ is studied using a data sample of $(448.1\pm2.9)\times10^{6}$ $\psi(3686)$ events~\cite{BESIII:2017tvm} collected by the BESIII detector, whose design and performance are described in Ref.~\cite{Ablikim:2009aa}. Candidate events are required to have at least four charged tracks in the main drift chamber (MDC) within the polar angle ($\theta$) range of $|\cos\theta| < 0.93$, where $\theta$ is defined with respect to the $z$-axis, which is the symmetry axis of the MDC. Monte Carlo (MC) studies show that charged tracks with momenta below 0.6 GeV$/c$ are dominated by pions, while those above 0.8 GeV$/c$ are mostly protons. The $\Lambda(\bar{\Lambda})$ candidates are reconstructed from $p\pi^{-}(\bar{p}\pi^{+})$ combinations. They are constrained to originate from a common vertex. The decay length ($L$) of the $\Lambda(\bar{\Lambda})$ candidates is required to be greater than twice the decay vertex resolution $(\sigma_L)$ away from the interaction point. If multiple $\Lambda\bar{\Lambda}$ pairs are found, the pair minimizing $(M_{p\pi^-}-m_{\Lambda})^2+(M_{\bar{p}\pi^+}-m_{\Lambda})^2$ is retained for further analysis, where $m_{\Lambda}$ is the nominal $\Lambda$ mass~\cite{ParticleDataGroup:2024cfk}. A four-constraint (4C) kinematic fit is applied to the $e^+e^-\to p\pi^+\bar{p}\pi^-$ hypothesis, constraining the total reconstructed four-momentum to that of the initial state. Events with fit quality $\chi^2_{\mathrm{4C}} < 100$ are retained. Additionally, the invariant masses of $p\pi^-$ and $\bar{p}\pi^+$ are required to be within the region [1.108, 1.123] GeV/$c^2$. After all selection criteria, 23,560 signal candidate events are selected for further analysis.

\begin{figure}[hbtp]
    \centering
    \includegraphics[width=0.65\linewidth,trim=0 8 15 10,angle=0]{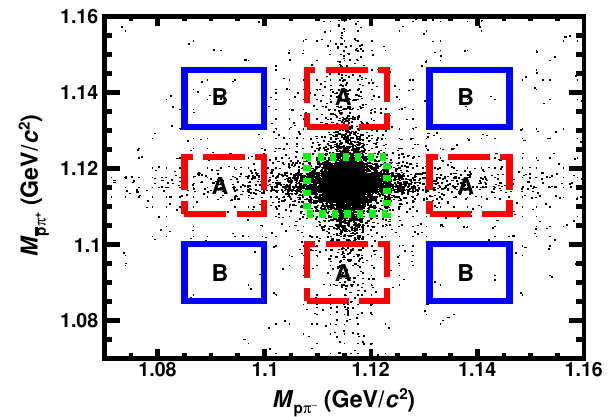}
\caption{Distribution of $p\pi^-$ versus $\bar{p}\pi^+$ for data. The green dotted box represents the $\Lambda \bar{\Lambda}$ signal region. The red dashed boxes indicate single-$\Lambda$ regions, while the blue solid boxes correspond to double-sideband regions.}
\label{fig2}
\end{figure}

After the above selection, MC studies~\cite{Zhou:2020ksj} indicate that the peaking backgrounds are found to be $\psi(3686)\to\Sigma^0\bar{\Sigma}^0$, $\psi(3686)\to\gamma\Lambda\bar{\Lambda}$, and $\psi(3686)\to\Sigma^0\bar{\Lambda} + \mathrm{c.c.}$, whose combined contributions are estimated to be 0.3\% of the total data sample. These contributions are found to be negligible. The numbers of non-$\Lambda\bar{\Lambda}$ background events, such as $\psi(3686)\to\pi^+\pi^- J/\psi\to\pi^+\pi^- p\bar{p}$ and $\psi(3686)\to\pi^+\pi^- p\bar{p}$, are estimated from the two-dimensional sideband regions in the data sample shown in Fig.~\ref{fig2}. The invariant mass sideband regions for $\Lambda$ and $\bar{\Lambda}$ are defined as [1.085, 1.100]~GeV/$c^2$ and [1.131, 1.146]~GeV/$c^2$. The number of background events $N_{\mathrm{bkg}}$ is determined by $0.5N_\mathrm{A}-0.25N_\mathrm{B}$, where $N_\mathrm{A}$ and $N_\mathrm{B}$ are the numbers of all events in the regions A and B, respectively. Using this method, the number of non-$\Lambda\bar{\Lambda}$ background events is estimated to be $653 \pm 18$. Additionally, only four events survive the event selection in the data sample collected at $\sqrt{s} = 3.65$~GeV. Using the formalism described in Eq.~(\ref{eq:anglW}) with the parameters $\alpha_{\psi}=0.82$, $\Delta\Phi=0.36$, and the Particle Data Group (PDG) values for $\alpha_{\Lambda}$ and $\alpha_{\bar{\Lambda}}$~\cite{ParticleDataGroup:2024cfk}, the number of continuum background events, scaled to the $\psi(3686)$ data, is estimated to be 60.

An unbinned maximum likelihood fit is performed on the five variables $(\theta_{\Lambda}, \theta_{p}, \phi_{p}, \theta_{\bar{p}}, \phi_{\bar{p}})$ to determine the parameters $\alpha_{\psi}$ and $\Delta\Phi$. The decay parameters ${\alpha_{\Lambda}}$(${\alpha_{\bar{\Lambda}}}$) are fixed to the PDG values~\cite{ParticleDataGroup:2024cfk}. The joint likelihood function is
\begin{equation}
\mathscr{L} = \prod_{i=1}^n Prob({\boldsymbol{\xi}}_i;{\boldsymbol{\Omega}}) = \prod_{i=1}^n \frac{{\mathcal{W}}({\boldsymbol{\xi_{i}}};
{\boldsymbol{\Omega}})}{\mathcal{N}},  
\end{equation}
where $Prob(\boldsymbol{\xi}_i)$ is the probability of observing event $i$ based on the observables ${\boldsymbol{\xi}}_i$ and the set of parameters $\boldsymbol{\Omega}$, $n$ is the number of events, $\mathcal{W}$ is calculated with Eq.~(\ref{eq:anglW}), and $\mathcal{N}$ is the normalization factor. The $\mathcal{N}$ is evaluated as $\frac{1}{N_\mathrm{MC}}\cdot\sum_{j=1}^{N_\mathrm{MC}}\mathcal{W}_j^\mathrm{MC}$, where $N_\mathrm{MC}$ is the total number of accepted MC events, $j$ labels the individual accepted MC event, and $\mathcal{W}_{j}^{\mathrm{MC}}$ is the angular distribution function evaluated for the $j$-th MC event. The MC events are generated according to the phase space model, undergo detector simulation, and are selected with the same event criteria as for data. In this way, the detection efficiency is naturally incorporated.
The parameters are determined using the MINUIT package from the CERN library~\cite{James:1975dr} to minimize the function defined as $S = -\mathrm{ln}\mathscr{L}_{\mathrm{data}} +\mathrm{ln}\mathscr{L}_{\mathrm{bkg}}$, where $\mathrm{ln}\mathscr{L}_{\mathrm{data}}$ and $\mathrm{ln}\mathscr{L}_{\mathrm{bkg}}$ are the log-likelihood functions for the data and the background events, respectively, with the latter including the non-$\Lambda\bar{\Lambda}$ background events and the continuum contribution.

\section{Systematic uncertainties}
The systematic uncertainties of the parameters $\alpha_{\psi}$ and $\Delta\Phi$ arise from the MC efficiency correction, event selection criteria, fit procedure, and background estimations. 
The uncertainties from tracking and decay length ratio are studied using the $J/\psi \to p\bar{p}\pi^+\pi^-$ and $J/\psi \to \Lambda\bar{\Lambda}$ control samples, respectively. The corresponding data-MC correction factors are applied to the nominal fit. To estimate the associated uncertainty, 500 MC phase space samples are generated where these factors are varied within one standard deviation of their statistical uncertainties. The fit results follow Gaussian distributions whose means match the measured values and whose widths are assigned as systematic uncertainties; the MC efficiency correction contributes 0.005 to $\alpha_{\psi}$ and $0.23^\circ$ to $\Delta\Phi$, while the decay length ratio correction contributes 0.000 and $0.06^\circ$, respectively. To estimate the kinematic fit uncertainty, track helix parameters ($\phi_{0}$, $\kappa$, and $\tan\lambda$) are corrected following Ref.~\cite{BESIII:2012mpj}. The difference between the results with and without these corrections is assigned as the uncertainty, which is 0.003 for $\alpha_{\psi}$ and $0.17^\circ$ for $\Delta\Phi$. 
The impact of the signal mass window is assessed by accounting for mass resolution and acceptance differences between data and MC simulation. Varying the window yields an uncertainty of 0.000 for $\alpha_{\psi}$ and $0.06^\circ$ for $\Delta\Phi$. 
To validate the fit results, 500 MC samples are generated using the differential cross-section in Eq.~(\ref{eq:anglW}) and the decay parameters from the nominal fit as inputs. Each sample has the same number of events as the data. The average output values are compared with the inputs, and their differences are taken as the systematic uncertainties from the fit method, amounting to 0.003 for $\alpha_{\psi}$ and $0.17^\circ$ for $\Delta\Phi$.
The uncertainties due to the fixed values of the decay asymmetries ($\alpha_{\Lambda}, \bar{\alpha}_{\Lambda}$) are estimated by varying them within their respective $\pm 1$ standard deviations, resulting in uncertainties of 0.005 for $\alpha_{\psi}$ and $0.34^\circ$ for $\Delta\Phi$.
The systematic uncertainty related to the fit range of $\cos\theta_{\Lambda}$ is estimated using an alternative fit range of $[-0.9, 0.9]$, and the resultant changes in the fit results are taken as the uncertainties, which are 0.001 for $\alpha_{\psi}$ and $0.06^\circ$ for $\Delta\Phi$.
The systematic uncertainty from the choice of the sideband regions is estimated by changing the ranges from [1.085, 1.100] GeV/$c^2$ and [1.131, 1.146] GeV/$c^2$ to [1.080, 1.095] GeV/$c^2$ and [1.136, 1.151] GeV/$c^2$. The background estimation and the fit are repeated, and the differences between the new and nominal fit results are taken as the systematic uncertainties, giving 0.001 for $\alpha_{\psi}$ and less than $0.01^\circ$ for $\Delta\Phi$.
For the continuum contribution, signal-like samples are generated with $\alpha_{\psi}=0$ and $\pm1$, while the continuum normalization is varied within the statistical uncertainty. The maximum deviation from the nominal fit is taken as the corresponding uncertainty, yielding 0.002 for $\alpha_{\psi}$ and $0.34^\circ$ for $\Delta\Phi$.
The potential bias in the parameters $\alpha_{\psi}$ and $\Delta\Phi$ due to the spin precession of the $\Lambda$ hyperon in the magnetic field is estimated to be less than 0.1\% based on the study in Ref.~\cite{Li:2019knv}, which is negligible compared to the statistical precision. By adding all these individual contributions in quadrature, the total systematic uncertainties for $\alpha_{\psi}$ and $\Delta\Phi$ are determined to be 0.009 and $0.59^\circ$, respectively.

\section{Results and Discussion}\label{sec:discussion} 
\begin{figure}[H]
\centering
\includegraphics[width=0.9\linewidth,trim=15 10 15 0,angle=0]{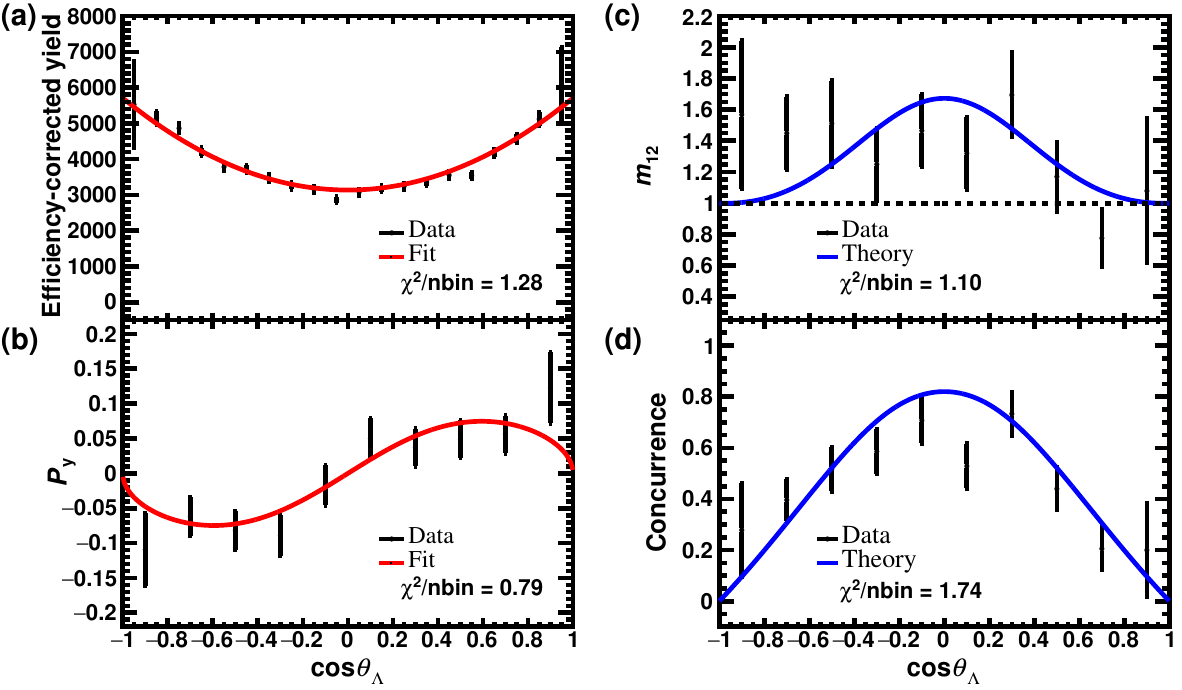}
\caption{Joint angular analysis and quantum correlation results as a function of $\cos\theta_{\Lambda}$: (a)~efficiency-corrected yield, with the red solid line represents the theoretical shape $1 + \alpha_{\psi}\cos^2\theta_\Lambda$ ($\alpha_{\psi} = 0.82$); (b)~$\Lambda$ transverse polarization $P_y$, with the red solid curve representing the expected angular dependence obtained with the parameters $\alpha_{\psi}$ and $\Delta\Phi$ determined from the fit to the full angular distribution; (c)~Bell nonlocality measure $m_{12}$ (dashed line indicating the CHSH bound of 1) and (d)~Wootters' concurrence $\mathcal{C}$, with the two blue solid curves representing the expectation based on the parameters $\alpha_{\psi}$ and $\Delta\Phi$. The black points represent the data, determined directly in (a) and via independent fits in (b)–(d). Error bars represent statistical uncertainties only in (a), and combined statistical and systematic uncertainties in (b)–(d). The $\chi^2/N_{\mathrm{nbin}}$ values denote the consistency between the data and the fits or expectations.}
\label{fig3}
\end{figure}

Based on $(448.1\pm2.9)\times 10^6$ $\psi(3686)$ events collected with the BESIII detector, the spin density matrix in the decay $\psi(3686) \to \Lambda\bar{\Lambda}$ is determined completely for the first time. 
The angular distribution parameter is determined to be $\alpha_{\psi} = 0.82 \pm 0.02_{\mathrm{stat.}} \pm 0.01_{\mathrm{syst.}}$, with a precision improved by a factor of more than three compared to the previous measurement~\cite{BESIII:2017kqw}, and a positive sign consistent with that $0.461\pm0.006_{\mathrm{stat.}}\pm0.007_{\mathrm{syst.}}$ from $J/\psi\to\Lambda\bar{\Lambda}$~\cite{BESIII:2018cnd}. This improvement arises from the multidimensional joint angular analysis, which exploits the full spin correlation of the $\Lambda \bar{\Lambda}$ system and thus makes more efficient use of the available angular information.
The relative phase is measured for the first time as $\Delta\Phi = (20.6 \pm 3.6_{\mathrm{stat.}} \pm 0.6_{\mathrm{syst.}})^{\circ}$.  The correlation coefficient between the fitted $\alpha_{\psi}$ and $\Delta\Phi$ is 0.330, indicating a positive correlation.
This non-zero phase implies a significant transverse $\Lambda$ polarization, with a maximum value of $P_{y} = 0.074 \pm 0.013_{\mathrm{stat.}} \pm 0.003_{\mathrm{syst.}}$ and a statistical significance of $6.3\sigma$ compared to the non-polarization hypothesis.  Figure~\ref{fig3}a shows the efficiency-corrected angular distribution of $\cos\theta_{\Lambda}$, where the data points are obtained by correcting the observed yields with the acceptance function projected onto the $\cos\theta_{\Lambda}$ axis. Figure~\ref{fig3}b displays the transverse polarization, where the $P_y$ data points are determined bin-by-bin using the same method as in Ref.~\cite{BESIII:2021ypr}. The independently determined data points in both panels agree well with the curves derived from the fit to the full angular distribution.

With the production parameters $\alpha_{\psi}$ and $\Delta\Phi$ determined from the global fit, independent fits are performed in each $\cos\theta_{\Lambda}$ bin within the same framework to extract the polarization $P_y$ and spin-correlation coefficients $C_{ij}$ ($i,j=x,y,z$), thereby providing a differential characterization of the joint spin state. 
Following Ref.~\cite{Wu:2024asu}, these observables are used to evaluate the Bell nonlocality measure $m_{12}$ and Wootters’ concurrence $\mathcal{C}$.
The Bell nonlocality measure is given by the sum of the two largest values in the set
\begin{equation}
C^{2}_{yy}, \frac{1}{4}\left[C_{xx} + C_{zz} \pm \sqrt{4C^{2}_{xz} +(C_{xx} - C_{zz})^{2}}\right]^2.
\label{eq:m12}
\end{equation}
The Wootters' concurrence is given by half of the maximum value of the set
\begin{equation}
\left\{0, \sqrt{4C_{xz}^2 + (C_{xx} - C_{zz})^2} - |1 - C_{yy}|, |C_{xx} + C_{zz}| - \sqrt{(1 + C_{yy})^2 - 4P_y^2}\right\}.
\end{equation}
Detailed formalism and additional results are provided in the Supplementary Material.
Figure~\ref{fig3}c and d present the Bell nonlocality measure $m_{12}$ and Wootters' concurrence $\mathcal{C}$ as functions of $\cos\theta_{\Lambda}$, respectively. The black points are obtained by inserting the independently fitted values of $P_y$ and $C_{ij}$ in each $\cos\theta_{\Lambda}$ bin into the above definitions. The blue curves represent the theoretical expectations obtained by first calculating $P_y$ and $C_{ij}$ from Eqs.~(\ref{eq:py})--(\ref{eq:cij}), using the values of $\alpha_{\psi}$ and $\Delta\Phi$ determined from the fit to the full angular distribution, and then evaluating $m_{12}$ and $\mathcal{C}$. Despite some local deviations, the data are overall consistent with the theoretical expectations within uncertainties. The concurrence $\mathcal{C}$ reaches its maximum at $\cos\theta_{\Lambda} = 0$, where its value equals $\alpha_{\psi}$. The average Bell nonlocality measure $\bar{m}_{12}$ is determined to be $1.266 \pm 0.078_{\mathrm{stat.}} \pm 0.049_{\mathrm{syst.}}$, obtained by taking the weighted average of the $m_{12}$ values from all angular bins, with weights given by the inverse squared measurement uncertainties. The significance is evaluated using a toy MC likelihood-ratio test based on the distributions of the test statistic under the Bell locality and Bell nonlocality hypotheses, yielding a CHSH inequality violation significance of $3.7\sigma$. This provides evidence for quantum nonlocality in the $\Lambda\bar{\Lambda}$ system produced in vector charmonium decays, within the same experimental framework that describes the production dynamics.

Compared with the results from $J/\psi \to \Lambda\bar{\Lambda}$ decay~\cite{BESIII:2018cnd}, the parameters $\alpha_{\psi}$ and $\Delta\Phi$  exhibit notable discrepancies. Using $\alpha_{\psi}$ and $\Delta\Phi$ within the covariant L-S framework~\cite{Wu:2021yfv}, we extract the ratio between the $D$- and $S$-wave coupling constants $g_{\mathrm{D}}/g_{\mathrm{S}}$, the relative phase $\delta$ between the $S$ and $D$ waves, the fraction of the $S$-wave contribution $\Gamma_{\mathrm{S}}/\Gamma_{\mathrm{total}}$, and the effective interaction radius $r_{\mathrm{eff}}$. These results are presented in Table~\ref{table1} and are compared with analogous parameters reported for other processes in Refs.~\cite{Wu:2021yfv,BESIII:2024nif}. The relative phase $\delta$ is found to be comparable for the $\Lambda\bar{\Lambda}$ production in both $J/\psi$ and $\psi(3686)$ decays, whereas for the $\Sigma^0\bar{\Sigma}^0$ production, $\delta$ differs by approximately $\pi$ between the $J/\psi$ and $\psi(3686)$ decays. The difference behavior of $\delta$ between the $\Lambda\bar{\Lambda}$ and $\Sigma^0\bar{\Sigma}^0$ final states presents an intriguing puzzle. A possible explanation lies in their different diquark structures: the $\Lambda$ contains a spin-singlet $ud$ diquark configuration, whereas the $\Sigma$ possesses a spin-triplet configuration. In the diquark model of Ref.~\cite{Jakob:1993th}, exact SU(6) spin-flavor symmetry gives $G^{\Sigma^{0}}_{\mathrm{M}} = -G^{\Lambda}_{\mathrm{M}}$, with an analogous relation for the Pauli form factor $F_2$. Such sign differences and momentum dependences may lead to different relative phases of the effective form factors for the $\Lambda$ and $\Sigma^{0}$, depending on the charmonium state and production dynamics. Although derived for space-like form factors, these relations provide qualitative insight into the contrasting polarization behavior at the $J/\psi$ and $\psi(3686)$ resonances and demonstrate the sensitivity of spin observables to internal quark configurations.

\begin{table}[h] 
\caption{The parameters of the $\psi\to\Lambda\bar{\Lambda}$ and  $\psi\to\Sigma^0\bar{\Sigma^0}$ processes, including the ratio of the $D$-wave and $S$-wave coupling constants $(g_{\mathrm{D}}/g_{\mathrm{S}})$, the relative phase between the $S$ and $D$ waves $(\delta)$, the fraction of the total width due to the $S$-wave $(\Gamma_{S}/\Gamma_{\mathrm{total}})$, and the effective radius $(r_{\mathrm{eff}})$.}
\label{table1}
\centering
\resizebox{\textwidth}{!}{%
\begin{tabular}{c c c c c}
\hline
Mode & $g_{\mathrm{D}}/g_{\mathrm{S}}$~(GeV$^{-2}$) & $\delta$~(rad) & $\Gamma_{\mathrm{S}}/\Gamma_{\mathrm{total}}$~(\%) & $r_{\mathrm{eff}}$~(fm)\\
\hline
$\psi(3686)\to\Lambda\bar{\Lambda}$ &$0.108\pm0.004\pm0.002$ & $-0.203\pm0.034\pm0.009$ & $83.7\pm1.0\pm0.4$ & $0.0438\pm0.0026\pm0.0011$\\
$J/\psi\to\Lambda\bar{\Lambda}$~\cite{Wu:2021yfv} & $0.180\pm0.005$ &$-0.804\pm0.024$ &$86.7\pm0.6$ & $0.0488\pm0.0021$ \\
$\psi(3770)\to\Lambda\bar{\Lambda}$ & $0.145\pm0.130\pm0.006$ &$-0.416\pm0.447\pm0.032$ &$71.3\pm29.5\pm1.6$ & $0.0740\pm0.0759\pm0.0040$ \\
$\psi(3686)\to\Sigma^0\bar{\Sigma}^0$~\cite{BESIII:2024nif} & $0.123\pm0.008\pm0.006$ &$-0.28\pm0.07\pm0.04$ &$82.7\pm1.9\pm1.4$ & $0.049\pm0.005\pm0.004$ \\
$J/\psi\to\Sigma^0\bar{\Sigma}^0$~\cite{BESIII:2024nif} & $0.1217\pm0.0015\pm0.0028$ &$2.947\pm0.017\pm0.012$ &$95.23\pm0.11\pm0.21$ & $0.0191\pm0.0005\pm0.0008$ \\
\hline 
\end{tabular}%
}
\begin{minipage}{0.95\textwidth}
\footnotesize
The first and second uncertainties are statistical and systematic, respectively, and the quoted single uncertainty represents the quadrature sum of these two components. The parameters for $\psi(3773)\to\Lambda\bar{\Lambda}$ are calculated in this work (the input values $\alpha_{\psi}$ and $\Delta\Phi$ from Ref.~\cite{BESIII:2021cvv}).
\end{minipage}
\end{table}

\section{Summary}\label{sec:summary} 

Using $(448.1\pm2.9)\times10^{6}$ $\psi(3686)$ events collected with the BESIII detector, we have achieved the first complete determination of the spin density matrix in $\psi(3686)\to\Lambda\bar{\Lambda}$. We measure $\alpha_{\psi}=0.82\pm0.02_{\mathrm{stat.}}\pm0.01_{\mathrm{syst.}}$ and determine the relative phase for the first time as $\Delta\Phi=(20.6\pm3.6_{\mathrm{stat.}}\pm0.6_{\mathrm{syst.}})^\circ$, resulting in a transverse polarization with a maximum value of $P_y=0.074\pm0.013_{\mathrm{stat.}}\pm0.003_{\mathrm{syst.}}$ and a significance of $6.3\sigma$. The complete spin density matrix further enables quantitative studies of spin correlations and quantum nonclassicality, yielding $\bar{m}_{12}=1.266\pm0.078_{\mathrm{stat.}}\pm0.049_{\mathrm{syst.}}$ and a significance of $3.7\sigma$ on CHSH inequality violation. Together with the L-S analysis, these results provide a unified experimental description of hyperon production dynamics and quantum correlations, and offer new insight for understanding the internal spin-flavor structure of hyperons.

\section*{Conflict of interest}
The authors declare that they have no conflict of interest.

\section*{Acknowledgments}
The BESIII Collaboration thanks the staff of BEPCII (https://cstr.cn/31109.02.BEPC) and the IHEP computing center for their strong support. This work is supported in part by National Key R\&D Program of China (2023YFA1606000, 2023YFA1606704); National Natural Science Foundation of China (NSFC) (12375070, 12247101, 11635010, 11935015, 11935016, 11935018, 12025502, 12035009, 12035013, 12061131003, 12192260, 12192261, 12192262, 12192263, 12192264, 12192265, 12221005, 12225509, 12235017, 12361141819); Chinese Academy of Sciences (CAS) Large-Scale Scientific Facility Program; Shanghai Top Talent Program of Eastern Talent Plan (BJZH2025073); Shanghai Leading Talent Program of Eastern Talent Plan (JLH5913002); Fundamental Research Funds for the Central Universities (lzujbky-2025-ytA05, lzujbky-2025-it06, lzujbky-2024-jdzx06); the Natural Science Foundation of Gansu Province (22JR5RA389, 25JRRA799); `111 Center' (B20063); the Strategic Priority Research Program of Chinese Academy of Sciences (XDA0480600); CAS (YSBR-101); 100 Talents Program of CAS; Institute of Nuclear and Particle Physics (INPAC) and Shanghai Key Laboratory for Particle Physics and Cosmology; European Research Council (ERC) (758462); German Research Foundation (DFG) (FOR5327); Istituto Nazionale di Fisica Nucleare, Italy; Knut and Alice Wallenberg Foundation (2021.0174, 2021.0299); Ministry of Development of Turkey (DPT2006K-120470); National Research Foundation of Korea (NRF-2022R1A2C1092335); National Science and Technology Fund of Mongolia; Polish National Science Centre (2024/53/B/ST2/00975); STFC (United Kingdom); Swedish Research Council (2019.04595); and U.S. Department of Energy (DE-FG02-05ER41374).



\begin{center}
	\textbf{BESIII Collaboration}
\end{center}
\begin{small}
M.~Ablikim$^{1}$\BESIIIorcid{0000-0002-3935-619X},
M.~N.~Achasov$^{4,b}$\BESIIIorcid{0000-0002-9400-8622},
P.~Adlarson$^{81}$\BESIIIorcid{0000-0001-6280-3851},
X.~C.~Ai$^{86}$\BESIIIorcid{0000-0003-3856-2415},
R.~Aliberti$^{38}$\BESIIIorcid{0000-0003-3500-4012},
A.~Amoroso$^{80A,80C}$\BESIIIorcid{0000-0002-3095-8610},
Q.~An$^{77,63,\dagger}$,
Y.~Bai$^{61}$\BESIIIorcid{0000-0001-6593-5665},
O.~Bakina$^{39}$\BESIIIorcid{0009-0005-0719-7461},
Y.~Ban$^{49,g}$\BESIIIorcid{0000-0002-1912-0374}
% ...
\\
\vspace{0.2cm}
(BESIII Collaboration)\\
\vspace{0.2cm} {\it
$^{1}$ Institute of High Energy Physics, Beijing 100049, People's Republic of China\\
% ...
$^{4}$ Budker Institute of Nuclear Physics SB RAS (BINP), Novosibirsk 630090, Russia\\
% ...
$^{38}$ Johannes Gutenberg University of Mainz, Johann-Joachim-Becher-Weg 45, D-55099 Mainz, Germany\\
$^{39}$ Joint Institute for Nuclear Research, 141980 Dubna, Moscow region, Russia\\
% ...
$^{49}$ Peking University, Beijing 100871, People's Republic of China\\
% ...
$^{61}$ Southeast University, Nanjing 211100, People's Republic of China\\
% ...
$^{63}$ State Key Laboratory of Particle Detection and Electronics, Beijing 100049, Hefei 230026, People's Republic of China\\
% ...
$^{77}$ University of Science and Technology of China, Hefei 230026, People's Republic of China\\
% ...
$^{80}$ University of Turin and INFN, (A)University of Turin, I-10125, Turin, Italy; (B)University of Eastern Piedmont, I-15121, Alessandria, Italy; (C)INFN, I-10125, Turin, Italy\\
$^{81}$ Uppsala University, Box 516, SE-75120 Uppsala, Sweden\\
% ...
$^{86}$ Zhengzhou University, Zhengzhou 450001, People's Republic of China\\

\vspace{0.2cm}
$^{\dagger}$ Deceased\\
% ...
$^{b}$ Also at the Novosibirsk State University, Novosibirsk, 630090, Russia\\
% ...
$^{g}$ Also at State Key Laboratory of Nuclear Physics and Technology, Peking University, Beijing 100871, People's Republic of China\\
% ...
}
\end{small}

\end{document}